\documentclass{aastex}
\usepackage{spr-astr-addons}
\usepackage{url}
\usepackage{graphicx}
\usepackage{epsfig}
\usepackage{epstopdf}
\usepackage{amsmath}

\RequirePackage{color}

\begin{document}

\title{Allowed and unique first-forbidden stellar electron emission rates of neutron-rich copper isotopes}
\shorttitle{Allowed and U1F rates of neutron-rich copper isotopes}
\shortauthors{Majid, Nabi and Daraz}

\author{Muhammad Majid\altaffilmark{1}} \author{Jameel-Un
Nabi\altaffilmark{1*}} \author{Gul Daraz\altaffilmark{1}}

\email{abc.com}

\altaffiltext{1}{Faculty of Engineering Sciences,\\GIK Institute of
Engineering Sciences and Technology, Topi 23640, Khyber Pakhtunkhwa,
Pakistan.} \altaffiltext{*}{Corresponding author email:
jameel@giki.edu.pk}

\begin{abstract}
The allowed charge-changing transitions are the most common weak
interaction processes of spin-isospin form that play a crucial role
in several nuclear/astrophysical processes. The first-forbidden (FF)
transition becomes important, in the circumstances where allowed
Gamow-Teller (GT) transitions are unfavored, specifically for
neutron-rich nuclei due to phase space considerations. In this paper
deformed proton-neutron quasi-particle random phase approximation
(pn-QRPA) model is applied, for the first time, for the estimation
of allowed GT and unique first-forbidden (U1F) transitions
($|\Delta$J$|$ = 2) of neutron rich copper isotopes in mass range 72
$\leq$ A $\leq$ 82 under stellar conditions. We compared our
computed terrestrial $\beta$-decay half-life values with previous
calculations and experimental results. It was concluded that the
pn-QRPA calculation is in good accordance with measured data. Our
study  suggests that the addition of rank (0 and 1) operators in FF
transitions can further improve the comparison which remain
unattended at this stage.  The deformed pn-QRPA model was employed
for the estimation of GT and U1F stellar electron emission
($\beta$$^{-}$-decay) rates over wide range of stellar temperature
(0.01 GK -- 30 GK) and density (10 -- 10$^{11}$ g/cm$^{3}$) domains
for astrophysical applications. Our study shows that, in high
density and low temperature regions, the contribution of U1F rates
to total electron emission rates of neutron-rich copper nuclei is
negligible.
\end{abstract}


\section{Introduction}
In this paper we calculate and discuss the stellar $\beta$-decay
processes of some of the astrophysically important neutron-rich
copper (Cu) isotopes. A precise understanding of the $\beta$-decay
properties of heavy nuclei is essential towards the understanding of
supernova explosion, especially for the comprehension of the
$r$-process nucleosynthesis. The elemental distribution on the
$r$-process path and the consequential final distribution of stable
nuclei are sensitive to the electron/positron emission
characteristics, particularly for heavier isotopes (neutron-rich)
taking part in these processes \citep{kla83,gro90}. Thousands of
nuclei are present between the neutron drip line and the line of
stability. In terrestrial laboratories majority of these nuclei
cannot be synthesized and necessitates the theoretical calculation
of nuclear structure and associated weak interaction properties for
these unstable nuclei. The availability of electron-neutrino
captures in the neutron-rich scenario greatly improves and amplifies
the $\beta$-decay effects. The successive neutrino-induced neutron
reactions also contribute by modifying the distribution shape of
$r$-abundance pattern \citep{mcl97}.

The nuclear weak interaction calculations of iron-regime and heavy
nuclei are believed to be  the key inputs that play a conclusive
role in determining various stellar processes. These include, but
are not limited to, the massive stars hydrostatic burning phases,
the presupernova evolution process of high mass stars and
nucleosynthesis processes like the $r$, $s$, $p$ and $rp$ -processes
\citep{bur57,wal81}. It is well known that the weak-decay stellar
reactions are governed by the Gamow-Teller (GT)  and, to a lesser
extent, by the Fermi transitions, for stellar densities $\rho$
$\lesssim$ $10^{11}$$g.cm^{-3}$. In addition, for nuclei possessing
chemical potential $\geq$ 30 MeV, situated in the region of
stability line, the forbidden weak interaction transitions also
contribute effectively for stellar densities $\rho$ $\gtrsim$
$10^{11}$$g.cm^{-3}$ \citep{coo84}. The electron capturing process
is one of the important weak-decay rates throughout the course of
presupernova stages of stellar evolution. During the shell burning
stages of silicon (Si), the electron emission (EE) rates cool down
the star and thereby compete with the electron capturing process.
During the Si burning phase, the EE rates are on the rise and this
increasing rate has consequences. The EE reactions are considered an
extra source for the neutrino production and this process cools the
stellar core and reduces the core entropy. Beyond the silicon shell
burning phases, the EE again contributes in cooling down the
temperature of the stellar core. As the stellar core stiffens to
high densities, the allowed weak-rates of EE become rather
insignificant. This is because of a substantial increase in the
electron Fermi energy which consequently chokes  the available phase
space. For the heavier (neutron-rich) nuclei the first-forbidden
(FF) transitions become significant as a larger phase space is
available for such transitions. For the neutron-rich nuclei, precise
and reliable estimations of $\beta$-decay half-lives (T$_{1/2}$) are
essential for a better understanding of late stages of stellar epoch
and the nucleosynthesis processes (specially the $r$-process).
Half-life estimates are  required for the purpose of experimental
investigation of nuclear properties and for designing purpose of
future radioactive ion-beam experimental facilities.

Due to the scarcity of measured data, use of theoretical models
became ever demanding for computation of weak-decay rates for
majority of the unstable nuclei. Several nuclear models have been
suggested and applied for the determination of the T$_{1/2}$ over
the past decades. Special mentions would include the Statistical
Gross Theory (SGT) \citep{tak73}, the pn-QRPA model
\citep{sta90,hir93} and shell model calculation \citep{Mut91} (the
last two models being microscopic in nature). Hybrid model, using
the pn-QRPA model for GT and SGT for FF decays, was developed by
\citep{Mol03} to perform the half-life calculation. Model
calculations were also performed in which the ground level of the
parent nucleus was determined by the Hartree-Fock-Bogoliubov (HFB)
scheme using Skyrme force in a fully self-consistent pn-QRPA
\citep{eng99}. \cite{Bor05} used the DF3 + CQRPA (continuum QRPA)
model and studied the contribution of allowed as well as FF
transitions to the total half-lives of $r$-process nuclei. It was
concluded that the FF contribution was indeed small to the total
half lives for A $\leq$ 78, but substantial for nuclei A $\geq 79$.
Under terrestrial conditions, the microscopic investigations,
employing the pn-QRPA model, for the allowed weak transitions were
performed by \citep{Kla84, sta90, hir93} and for the unique
first-forbidden (U1F) transitions by \citep{Hom96}. More recently
\citep{Zhi13} used the large-scale shell-model (LSSM) and computed
the allowed and FF contribution to the total T$_{1/2}$ of the
$r$-process waiting point nuclei. \cite{Bor05} used the CQRPA model
for the prediction of beta decay half-lives (GT + U1F) of
neutron-rich isotopes. However the Borzov results were limited only
to spherical nuclei. Recently, authors in \cite{Hos10} measured the
T$_{1/2}$ and branching ratios for neutron-rich nuclei and found
that the deformation of nucleus can have a considerable effect on
the T$_{1/2}$ and on $\beta$-delayed neutron emission probabilities.
Further to this there was a need to extend these calculations to
finite temperature domain, in a microscopic fashion, in order to
better comprehend the working mechanism of numerous astrophysical
processes. Two such widely used approaches are the shell model
\citep{Lan00} approach and the pn-QRPA model \citep{nab99} approach.
Stellar weak rates calculated using the two models were used for
numerous astrophysical applications. The pn-QRPA model presents a
state by state evaluation of the weak charge-changing transitions.
GT strength distributions are calculated microscopically from parent
excited states. In other words this model is the only model that
does not employ the Brink-Axel hypothesis \citep{Bri55} in stellar
weak rate calculations.

In this paper we use the deformed pn-QRPA model for the prediction
of T$_{1/2}$ of neutron-rich copper isotopes having mass range 72
$\leq$ A $\leq$ 82. For these isotopes previous theoretical and
experimental half-lives are available in literature.  Employing the
same model, we calculate the allowed GT and U1F weak interaction EE
rates in stellar matter for these neutron-rich Cu isotopes. As
mentioned earlier our calculation of stellar weak rates do not
assume Brink-Axel hypothesis as adopted in previous calculations.
This makes the current calculation unique and fully microscopic in
nature. Motivation of the present calculations also came, in part,
from the work of \citep{Hom96}. There the authors pointed out that
the U1F transitions have a large contribution to the total
$\beta$-decay weak rates. Homma and collaborators  got better
results, for their calculated T$_{1/2}$, using the deformed pn-QRPA
theory \citep{Hom96}. However the non-unique FF transition
contributions (rank 0 and rank 1) are also important and currently
missing in our calculation. We plan to calculate non-unique FF
transitions in future.

The  formalism for the computation of allowed GT and U1F weak
interaction rates using deformed pn-QRPA will be discussed in
Sec.~2. We present our calculated results in Sec.~3. Here we also
compare our result with measured and previous theoretical
calculations. We conclude in Sec.~4 by summarizing the main features
of this work.

\section{Theoretical Formalism}
The description for the calculation of allowed and FF transitions
are given in \citep{sch66, wu66,com73,Bla79, beh82}. It is rather
simple to calculate the allowed charge-changing transitions. The FF
transitions, on the other hand, allow a wide spectrum in nuclear
matrix elements as well as in lepton kinematics
\citep{wu66,sch66,beh82}.

In our model a quasi-particle basis was constructed (using the
Bogoliubov transformation). Pairing correlations between nucleons
were treated within the BCS approximation (using a constant pairing
force). The random phase approximation equation was solved with the
GT residual interaction of schematic separable forces. Deformed
Nilsson potential oscillator, along with a quadratic deformation,
was used for the estimation of single-particle energies and wave
functions. In stellar scenario, the pn-QRPA model formalism, used to
estimate the allowed weak interaction rates may be seen in detail
from Refs. \citep{nab99, nab04}. Below we briefly describe the
pn-QRPA formalism used for the estimation of U1F electron emission
(EE) rates.

The calculation for the EE rates of U1F transitions were carried out
using separable forces in the nuclear matrix elements, which appear
in the usual RPA equation,
\begin{equation}
\label{vph} V^{ph}_{pn,p^{\prime}n^{\prime}} = +2\chi
q_{pn}(\mu)q_{p^{\prime}n^{\prime}}(\mu),
\end{equation}

\begin{equation}
\label{vpp} V^{pp}_{pn,p^{\prime}n^{\prime}} = -2\kappa
q_{pn}(\mu)q_{p^{\prime}n^{\prime}}(\mu),
\end{equation}
where
\begin{equation}
\label{fpn} q_{pn}(\mu)=<j_{p}m_{p}|t_{-}r[\sigma
Y_{1}]_{2\mu}|j_{n}m_{n}>,
\end{equation}
are the transition amplitude for the single particle  U1F EE
transitions. Here  $m_{p(n)}$ represents third component of the
angular momentum for protons (neutrons) and $\mu = m_{p}-m_{n}$, the
possible values of $\mu$ are $\mu = 0, \pm1$, and $\pm2$. The
symbols $\chi$ and $\kappa$ characterize the strength of GT force in
particle-hole and particle-particle channel, respectively. Other
symbols have their usual meaning in the RPA formalism. In our
earlier calculations we deduced an $A$-dependent value for the
interaction constants ($\chi$ and $\kappa$) \cite{Nab15, Nab17}. The
value of $\chi$ adopted in our calculation is 61.20/A (for both
allowed GT and U1F transitions) whereas $\kappa =$  4.85/A (MeV) and
10.92/A (MeV fm$^{-2}$)) for the GT and U1F transitions,
respectively. The chosen values of $\chi$ and $\kappa$ display a
$1/A$ dependence as suggested in \citep{Hom96, Nab15, Nab16, Nab17}.
The deformation parameter $\beta_{2}$ for the selected copper
isotopes were adopted from \citep{Mol95}. $Q$-values were taken from
\cite{Aud12}.

The stellar EE rate for the U1F transitions from the
$\mathit{p^{th}}$ level of the parent  to the $\mathit{d^{th}}$
level of the daughter nucleus is given by

\begin{equation}
\label{lij} \lambda_{pd}^{EE} =
\frac{m_{e}^{5}c^{4}}{2\pi^{3}\hbar^{7}}\sum_{\Delta
J^{\pi}}g^{2}f_{pd}(\Delta J^{\pi})B_{pd}(\Delta J^{\pi}),
\end{equation}
where $ (\Delta J^{\pi} = 2^{-})$. In Eq.~\ref{lij}, $ B_{pd}(\Delta
J^{\pi})$ are the reduced transition probabilities and are
associated with the nuclear matrix elements of the U1F weak
interaction, $f_{pd}(\Delta J^{\pi})$ are the phase-space integrals
and $g$ represents the weak coupling constant. The other constants
have their usual meaning.

The phase-space factors $f_{pd}(\Delta J^{\pi})$ are taken as
integrals over the lepton distribution functions and therefore are
sensitive to the temperature and density of the stellar medium.
Using natural units ($\hbar = c = m_e =1$), the integrals are given
by
\begin{eqnarray}
f_{pd} = \int_{1}^{w_{m}} w \sqrt{w^{2}-1}
(w_{m}-w)^{2}[(w_{m}-w)^{2}F_{1}(Z,w) \nonumber\\
+ (w^{2}-1)F_{2}(Z,w)] (1-G_{-}) dw,
\end{eqnarray}
where $w$ is the total kinetic energy  of the electron including its
rest mass and $w_{m}$ gives the total $\beta$-decay energy ($ w_{m}
= m_{p}-m_{d}+E_{p}-E_{d}$, where $m_{p}$ and $E_{p}$ are the mass
and excitation energies of the parent nucleus, and $m_{d}$ and
$E_{d}$ of the daughter nucleus, respectively). $G_{-}$ is the
electron distribution function. Our model calculation assumes that
the electrons are not in the bound state and then $G_{-}$  is simply
the Fermi-Dirac distribution function
\begin{equation}
G_{-} = [exp(\frac{E-E_{f}}{kT})+1]^{-1},
\end{equation}
here $E=(w-1)$ is the kinetic energy of the electrons, $E_{f}$ is
the Fermi energy of the electrons, $T$ is the temperature, and $k$
is the Boltzmann constant. The Fermi functions, $F_{1}(\pm Z,w)$ and
$F_{2}(\pm Z,w)$ appearing in Eq.~(5), were computed according to
the procedure adopted by \citep{Gov71}. The electron  number density
related with nuclei and protons is $\rho Y_{e} N_{A}$, where $\rho$
represent the baryon density, $Y_{e}$ is the ratio of electron
number to the baryon number, and $N_{A}$ is the Avogadro's number.
\begin{equation} \rho Y_{e} = \frac{1}{\pi^{2}N_{A}}(\frac
{m_{e}c}{\hbar})^{3} \int_{0}^{\infty} (G_{-}-G_{+}) p^{2}dp,
\end{equation}
here $p=(w^{2}-1)^{1/2}$ gives the positron or electron momentum,
 $G_{+}$ is the positron distribution
function given by
\begin{equation} G_{+} =\left[\exp
\left(\frac{E+2+E_{f} }{kT}\right)+1\right]^{-1}.
\end{equation}
Eq.~(7) is used for an iterative calculation of Fermi energies at
chosen values of $T$ and $\rho Y_{e}$.

In the stellar scenario the temperature of the  core of massive
stars may be high enough facilitating a finite probability of
occupancy of parent excited levels. The contribution to the total
weak rates from these excited levels  may  be significant. For a
state $p$, the probability of occupancy can be calculated using the
Boltzmann thermal equilibrium assumption,

\begin{equation}
\label{pi} P_{p} = \frac {exp(-E_{p}/kT)}{\sum exp(-E_{p}/kT)},
\end{equation}
where $E_{p}$ shows the excitation energy of the $p^{th}$ parent
level.

During the weak interaction process the electron emission rate per
nucleus per unit time can be calculated using,
\begin{equation}
\label{lb} \lambda^{EE} = \sum_{pd}P_{p} \lambda_{pd}^{EE}.
\end{equation}
Many discrete states can contribute to the two sums over parent and
daughter energy eigenvalues. The core temperature makes the thermal
population of excited states in the parent possible. Each of these
parent states may be connected to many levels in the daughter
nucleus via the GT operator. In this work the summation was applied
on 150 parent and same number of daughter excited states to compute
the total EE rates. The summation was applied and satisfactory
convergence was ensured for the total weak-decay rate. Good examples
on convergence of stellar weak rates can be seen in Ref.
\citep{Oda94} (see their discussion on Table A) and in Ref.
\citep{Lan00} (see their discussion on Tables 1 and 2). It may be
noted that the availability of multi-$\hbar\omega$ model space
(7$\hbar\omega$ oscillator space) in the deformed pn-QRPA model
makes it possible to perform a microscopic calculation of GT + U1F
strength distribution functions for $E_{p}$ well in excess of 10 MeV
without assuming Brink-Axel hypothesis.

\section{Results and Discussion}
During presupernova evolution of high mass stars, weak rates on
copper (Cu) isotopes are considered to play a crucial role. Previous
simulation result \citep{Auf94} shows that $^{72}$Cu, $^{74}$Cu and
$^{76}$Cu significantly alter the electron to-baryon ratio inside
the core of massive stars. In this section we report on the
calculation of terrestrial half-lives  (GT + U1F) and total electron
emission (EE) rates (GT +U1F) for heavy Cu isotopes in stellar
environment using the deformed pn-QRPA model.

The pn-QRPA computed terrestrial half-lives for our chosen Cu
isotopes are shown in Table~1. To highlight the improvement brought
by inclusion of U1F transitions, we show both the GT and (GT + U1F)
calculations of half-lives in Table~1.  We further compare our
results with previous theoretical and measured half-lives in
Table~1. Column~2 shows the T$_{1/2}$ (only GT) values of
\citep{Mol97} calculation based on FRDM + RPA model. In the FRDM +
RPA calculation, odd-even effects are prominent in the $\beta$-decay
T$_{1/2}$ values due to the exclusion of the useful \emph{pn}
interaction in particle-particle (pp) channel. Columns~3~-~5 show
the GT calculation by \citep{Pfe02}. Authors in \citep{Pfe02} used
the empirical Kratz Herrmann formula (referred to as KHF) and a
unified macroscopic-microscopic QRPA for their calculations. The
QRPA calculation was further divided into two category: QRPA-1 and
QRPA-2. For details of the KHF, QRPA-1 and QRPA-2 calculations we
refer to their paper. Column~6 shows the theoretical (GT + FF)
results of \citep{Bor05} calculations based on the (DF3 + CQRPA)
model. It is to be noted that Borzov calculation is limited only to
spherical nuclei. Recently \cite{Hos10} pointed out that the
spherical shape assumption is not desirable and deformation of the
nucleus can have a considerable effect on the $\beta$-decay
half-lives.  The last three columns list the experimental T$_{1/2}$
values from \citep{Aud12, Aud03} and \citep{Hos10}, respectively. It
is noted that the addition of U1F transition in our pn-QRPA
calculation improves the overall comparison with the measured data.
This comparison may still be improved by incorporation of non-unique
first-forbidden transitions. For now we are not in a position to
calculate these non-unique (rank 0 and 1) contributions but plan to
take this as a future assignment.

Recently the large-scale shell-model (LSSM) half-life results
including the FF transitions (having both unique and non-unique
contributions) were accomplished for waiting point nuclei possessing
magic neutron numbers $N$ = 50, 82 and 126 by \citep{Zhi13}. The
authors concluded that for nuclei having $Z \ge$ 28 and $N$ = 50,
the rank 2 operators contribute significantly to the FF transitions.
Table~2 shows the comparison of LSSM calculation with our pn-QRPA
and measured half-lives for $^{79}$Cu. It can be seen that LSSM
overestimates the calculated half-life. The pn-QRPA calculated
half-life (GT + U1F)  is in better agreement  with the measured
half-life \citep{Aud12}. \cite{Zhi13} used the quenching factor of
0.66 for all nuclei in case of allowed transitions. However, for
various nuclear matrix elements, in case of first-forbidden
transitions different quenching factors varying from 0.38 to 1.266
were employed in LSSM calculation. In our deformed pn-QRPA model a
quenching factor of 0.6 was used both for allowed and U1F
transitions. \citep{Vet89} and \citep{Ron93} predicted the same
quenching factor of 0.6 for the RPA calculation in the case of
$^{54}$Fe when comparing their measured strengths to RPA
calculations. The same quenching factor of 0.6 was also used by
\citep{Bro85, Mut91, Oda94}

In order to increase the reliability of our computed weak-decay
rates, we did incorporate measured energy levels (XUNDL) in our
calculation. Calculated excitation energies were replaced with the
measured states when they were within 0.5 MeV of each another.
Missing measured levels were added together with their
$\log$\emph{ft} values wherever appropriate. Calculated states were
not replaced beyond measured levels for which parity and/or spin
assignments were not certain. Experimental low-lying states were
inserted for $^{72-75,77}$Cu (allowed GT transitions). It is
pertinent to mention that previous theoretical computations, to
which we compared our results, did not incorporate measured data in
their calculations. Table~3 shows the difference in calculated
allowed EE rates with and without insertion of experimental data. It
is noted that, for low densities, the rates with experimental
insertion are substantially smaller than those calculated by our
model. This difference is caused primarily because of missing of few
low-lying states in our model calculation. However for T$_{9} \ge$
10 (typical core temperatures) the two calculations are in decent
comparison. T$_{9}$ signifies stellar temperature in units of
10$^{9}$ K. At high stellar densities and low temperatures the
allowed EE rates are negligible. Here we show the comparison once
the EE rates exceed 10$^{-12}$ s$^{-1}$. A fair comparison is again
noted.

The calculated allowed and U1F charge-changing transitions for
selected Cu isotopes are shown in Figs.~1-3. Calculated transition
strengths for $^{72,73}$Cu are shown in Fig.~1. Figs.~2 and 3 depict
the corresponding result for $^{76,77}$Cu and $^{81,82}$Cu,
respectively. The abscissa shows the daughter excitation energy in
MeV units. It should be noted that the calculated strength values
less than 10$^{-4}$ units, though calculated, are not displayed in
these figures. Figs.~1-3 show that the strength distributions
calculated using the deformed pn-QRPA model are well fragmented. It
can also be seen that U1F transition strength is significant for
selected copper isotopes and contributes in lowering of the
calculated half-lives as against those computed only from GT
transitions.

We employed the deformed pn-QRPA model, for the first time, to
compute allowed (GT) \textit{and} U1F electron emission (EE) rates
in stellar matter for neutron-rich Cu isotopes. The EE rates were
computed at stellar density in the range (10 -- 10$^{11}$)
g/cm$^{3}$ and at temperature ranging from T$_{9} $ = 0.01 -- 30.
Here we would like to discuss the high-temperature correction
effects as discussed by \citep{Rau03}. The maximum excitation energy
above which there are no more significant contributions to the
nuclear partition function is of the order of (20-25) MeV up to
T$_{9} \sim $ 10 \citep{Rau00}. The high-temperature corrections are
applicable at stellar temperatures T$_{9} $ = 50 -- 60 for light and
intermediate nuclei (considered in this paper) and as low as T$_{9}$
= 14 for heavy nuclei \citep{Rau03}. We did not incorporate any
high-temperature corrections in our current work primarily because
(i) we are not dealing with heavy nuclei, and, (ii) our temperature
range does not exceed T$_{9}$ = 30. However we note that
incorporation of high-temperature effects can lead to further
improvement in our calculated rates. In future we would like to work
on incorporation of high-temperature effects in our rate
calculation.

The allowed GT and U1F rates for selected Cu isotopes are shown in
Tables~4--6. The first column shows the stellar density
($\rho$Y$_{e}$) (in units of gcm$^{-3}$).  The calculated emission
rates are tabulated in logarithmic (to base 10) scale in units of
s$^{-1}$. These tables show that as the temperature rises, the EE
rates increase because of the finite contribution of excited state
partial rates. As the stellar core stiffens to high density, the EE
rates decrease substantially by orders of magnitude. High stellar
density reduces the available phase space for electrons because of
Pauli blocking. It can be seen from these tables that the
contribution of U1F rates is significant for heavier isotopes. Our
findings are in line with the conclusion of \citep{Bor05} that FF
contribution becomes significant for nuclei with $A \ge$ 79.

The contribution of  GT and U1F rates to total EE rate as a function
of stellar temperature  is shown in Figs.~4--6. Fig.~4 is drawn at
low stellar density ($\rho$Y$_{e}$) of 10$^{3}$ g/cm$^{3}$. Fig.~5
is drawn at $\rho$Y$_{e}$ = 10$^{7}$ g/cm$^{3}$ (intermediate
density) and Fig.~6 displays the results at $\rho$Y$_{e}$ =
10$^{11}$ g/cm$^{3}$ (high density). Each figure consist of six
panels illustrating the respective contribution of GT and U1F rates
at different stellar temperatures. Our findings show that the
contribution of U1F rates decreases with increasing stellar density.
It can also be seen that the U1F contribution is significant for
$^{80-82}$Cu. This confirms the conclusion of \cite{Bor05} that the
contribution of first-forbidden transitions is minor for nuclei with
mass number greater and/or equal to 78 due to smaller phase space
consideration. We note that for stellar temperatures T$_{9} < $ 15
 and high density regions, almost all contribution to total EE
rates comes from GT transitions. Tables~7--9 present the calculated
allowed and U1F rates, along with percentage GT contribution, for
all copper isotopes as a function of stellar temperature and
density.  EE rates (GT + U1F) on a fine-grid temperature-density
scale, suitable for interpolation purposes, are available as ASCII
files and may be requested from the authors.

\section{Conclusion}
The weak rates play a decisive role in the presupernova evolution of
high mass stars thereby controlling the electron to-baryon content
of the core. A fully microscopic calculation of weak rates (one
avoiding use of Brink-Axel hypothesis) is desirable in order to
obtain a realistic picture of astrophysical processes. In this paper
the deformed pn-QRPA model was used for the computation of
charge-changing transitions (GT + U1F), terrestrial half-lives and
electron emission rates  for neutron-rich copper nuclei in mass
range 72 $\leq$ A $\leq$ 82. Our deformed pn-QRPA calculated
half-lives, including both GT and U1F contribution, were in  good
agreement with available experimental data. We calculated EE rates
(GT + U1F) over a wide range of temperature (T$_{9}$ = 0.01-- 30)
and density (10--10$^{11}$ g/cm$^{3}$) scale. For $^{80-82}$Cu there
is a substantial contribution from U1F transitions to the total EE
rates in line with the findings made by Borzov. The U1F contribution
to total EE rates, however, decreases with increasing stellar
density.

\acknowledgments  J.-U. Nabi would like to acknowledge the support
of the Higher Education Commission Pakistan through the HEC Projects
No. 20-3099 and 20-5557.

\nocite{*}
\bibliographystyle{spr-mp-nameyear-cnd}
\bibliography{myref}
\bibliography{biblio-u1}

\begin{figure*}[]
\begin{center}
  \begin{tabular}{cc}
    \includegraphics[scale=0.35]{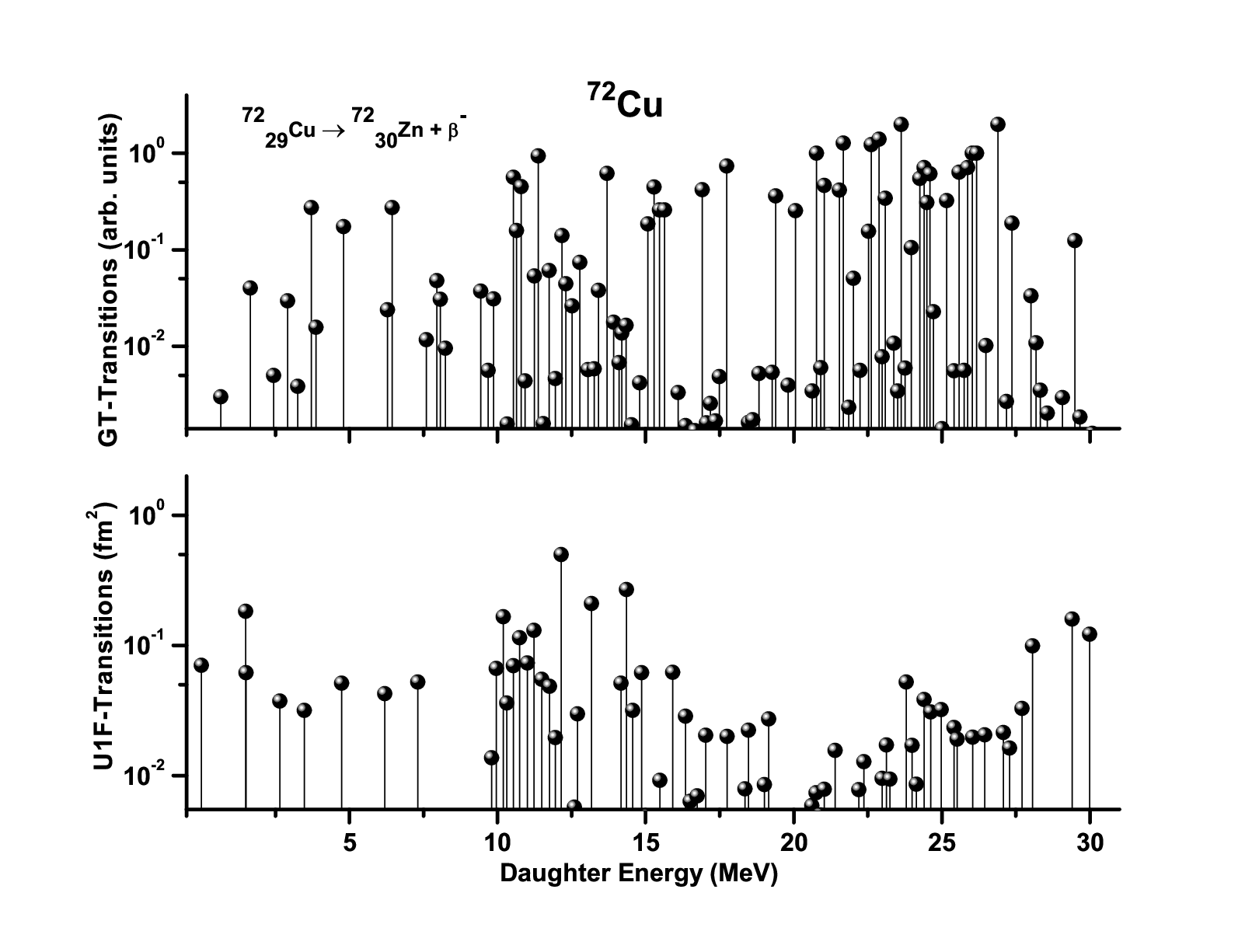} &
    \includegraphics[scale=0.35]{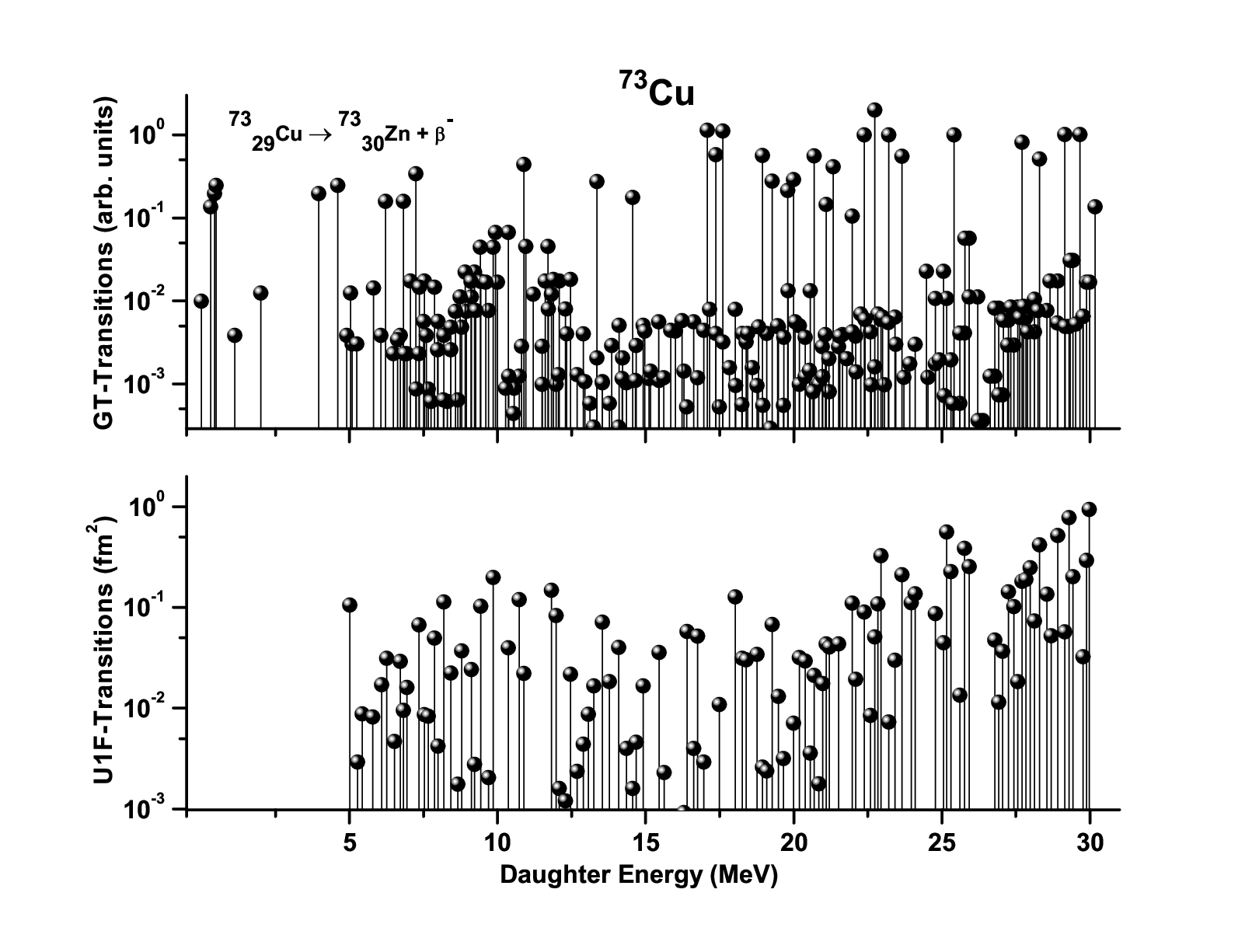}\\

\end{tabular}

\caption{The pn-QRPA calculated allowed and U1F transitions for
$^{72,73}$Cu isotopes in $\beta$-decay direction.}\label{fig1}
\end{center}
\end{figure*}

\begin{figure*}[]
\begin{center}
  \begin{tabular}{cc}
    \includegraphics[scale=0.35]{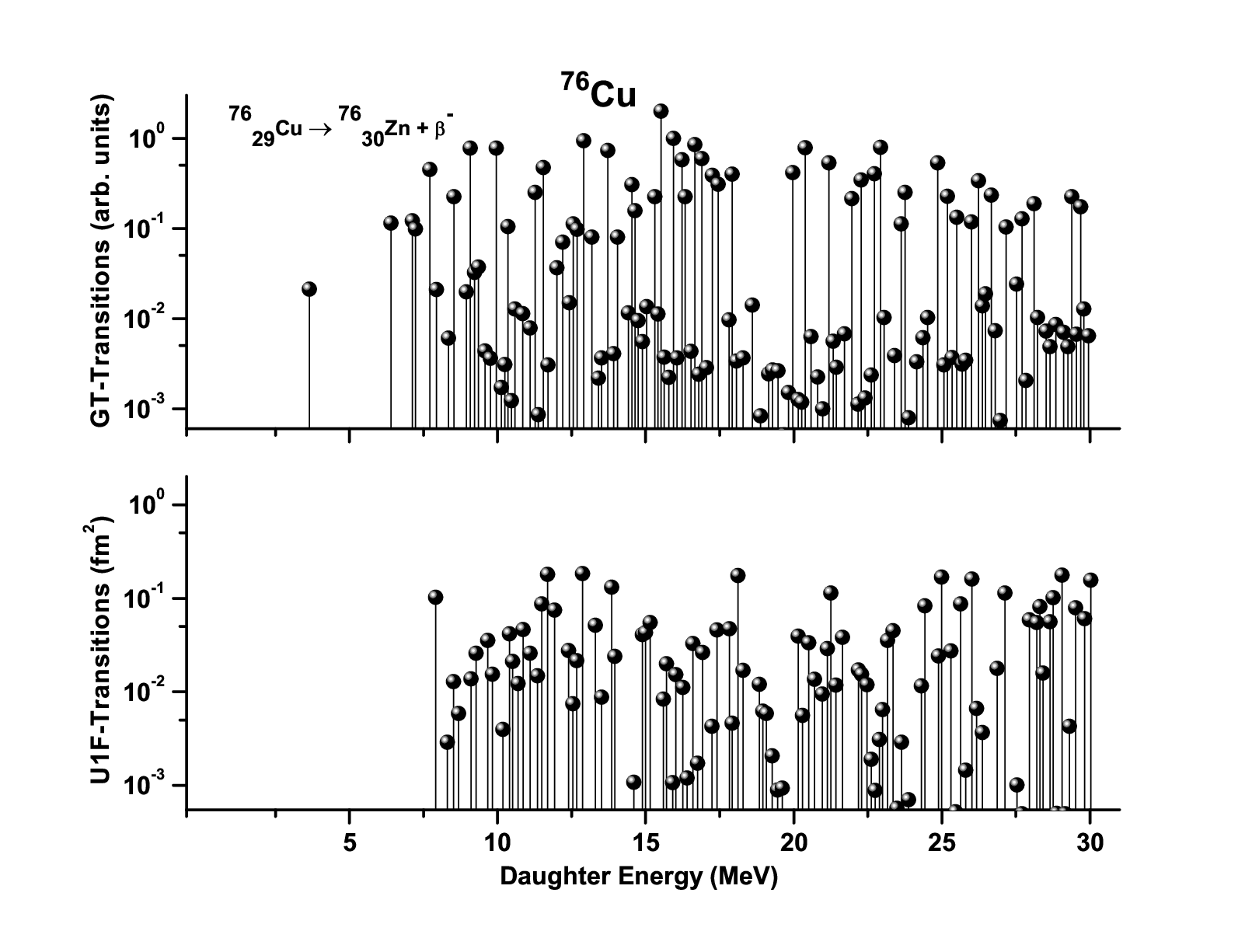} &
    \includegraphics[scale=0.35]{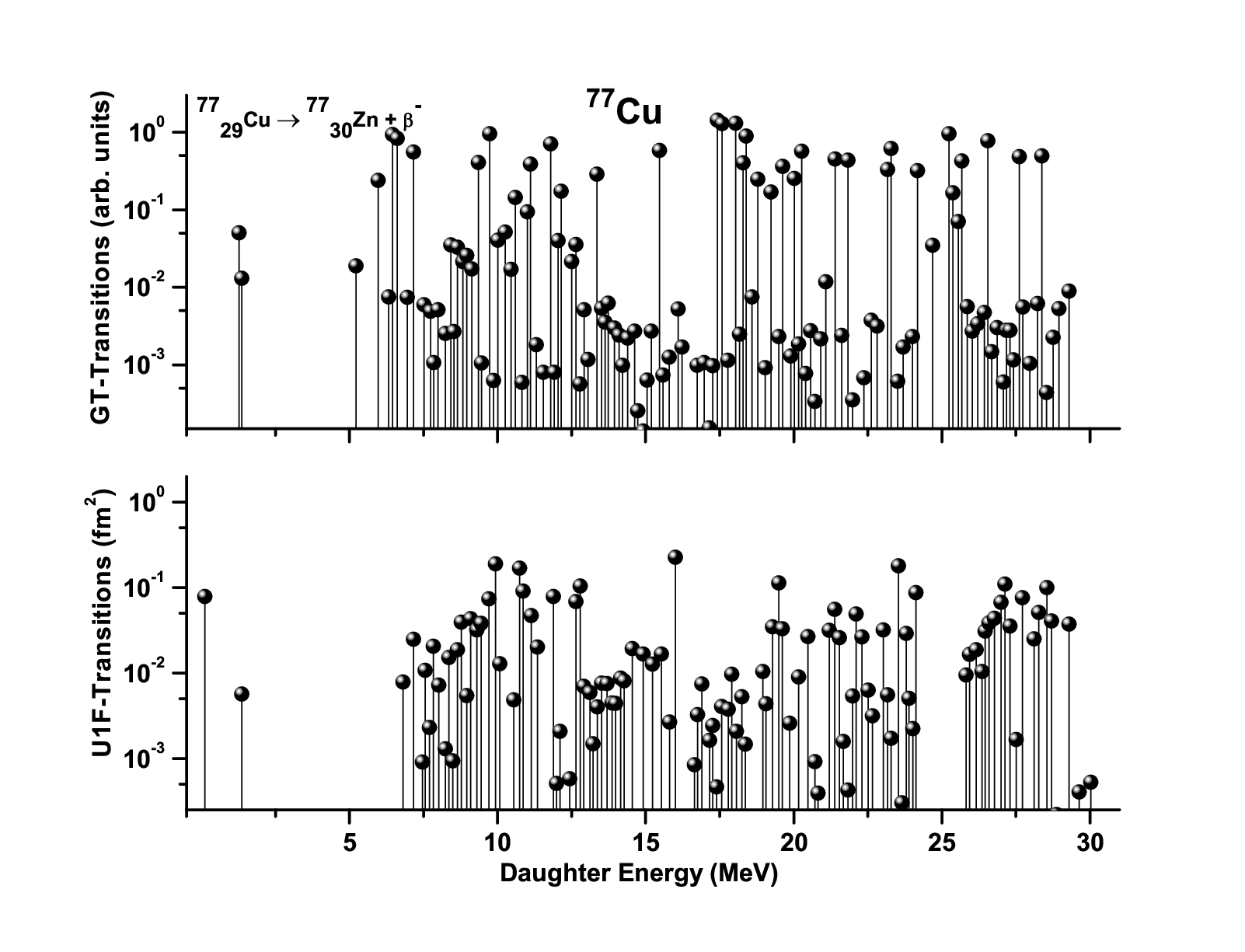}\\

\end{tabular}

\caption{Same as Fig. 1, but for $^{76,77}$Cu isotopes.}\label{fig2}
\end{center}
\end{figure*}

\begin{figure*}[]
\begin{center}
  \begin{tabular}{cc}
    \includegraphics[scale=0.35]{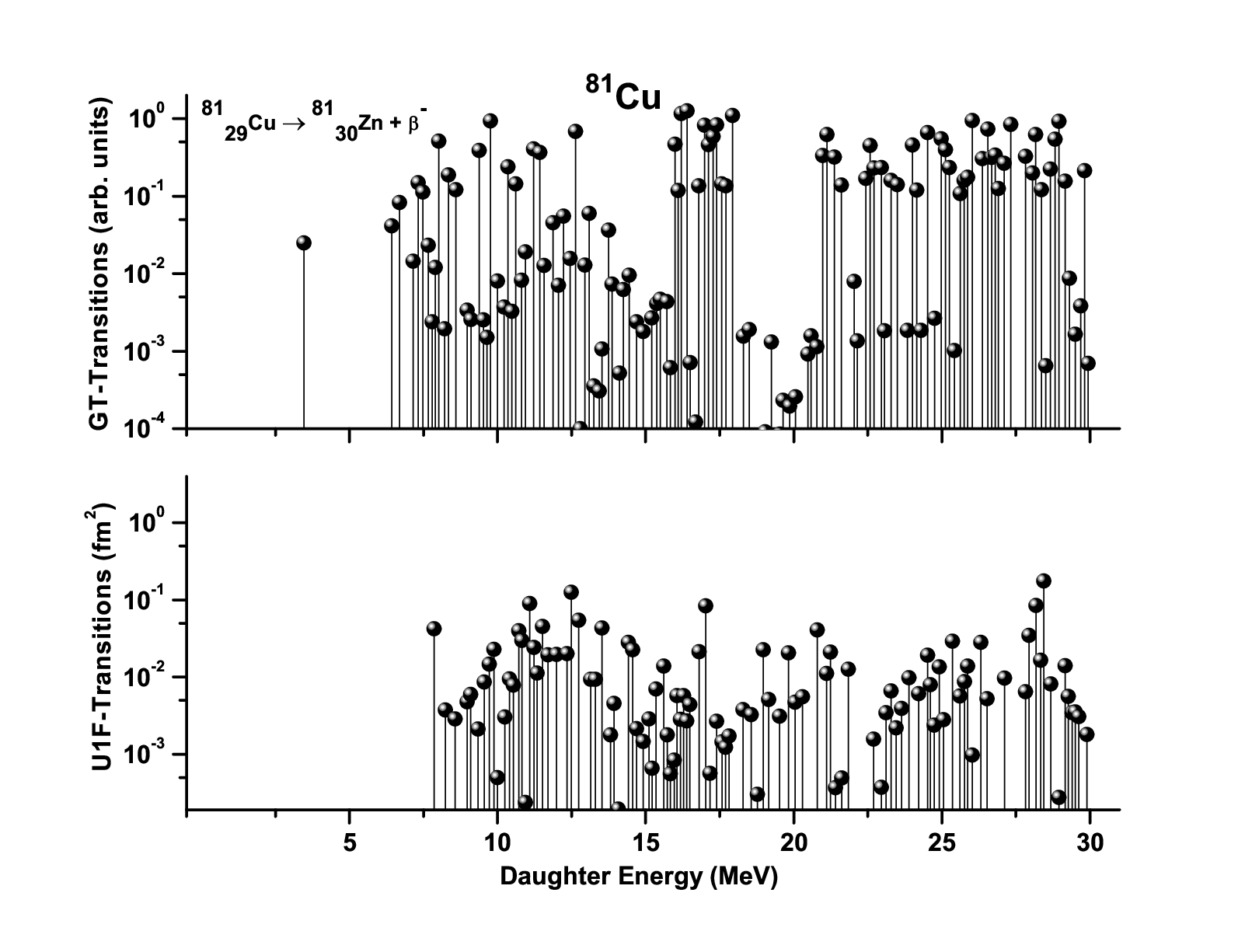} &
    \includegraphics[scale=0.35]{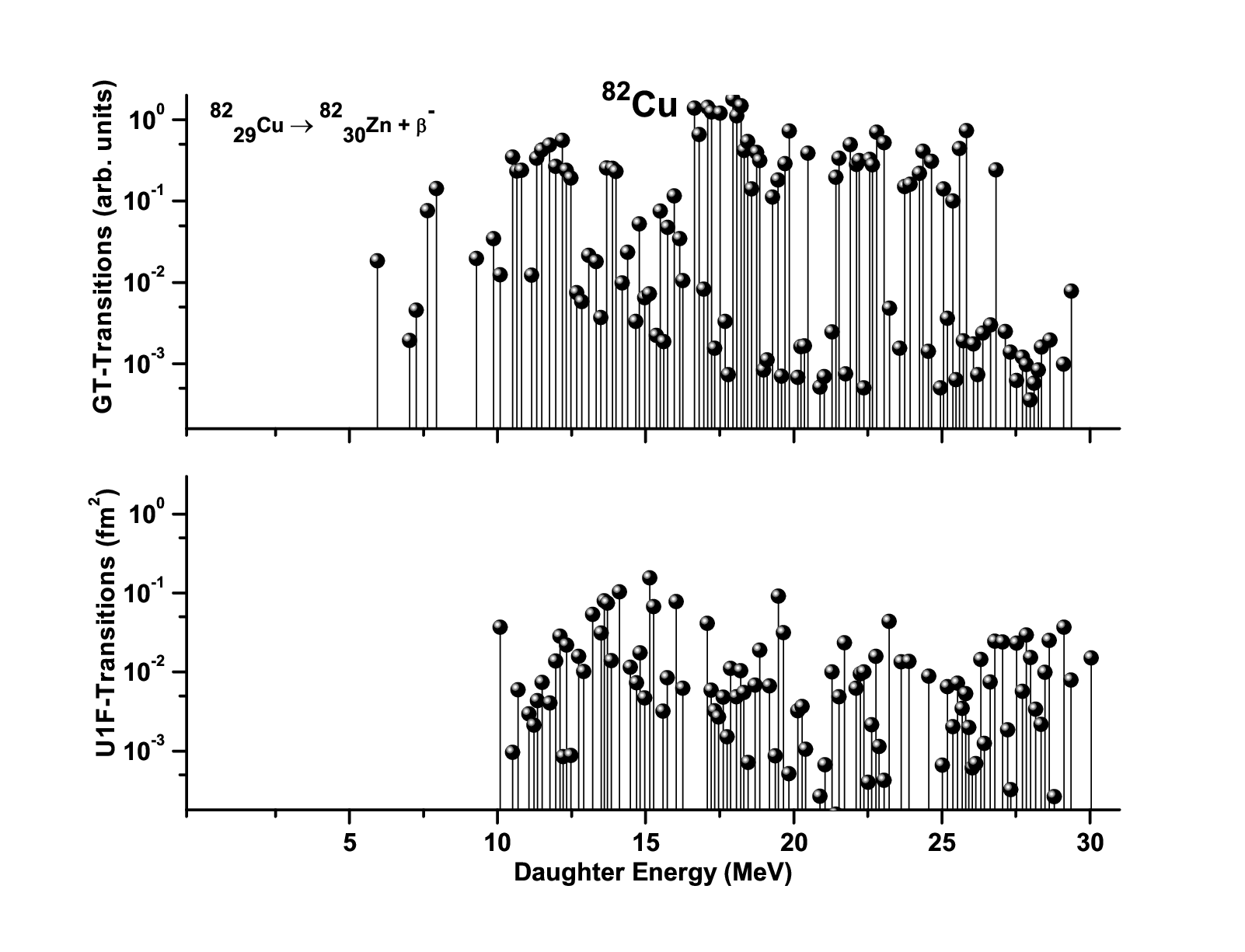}\\

\end{tabular}

\caption{Same as Fig. 1, but for $^{81,82}$Cu isotopes.}\label{fig3}
\end{center}
\end{figure*}

\begin{figure*}[h]
\includegraphics[scale=0.6]{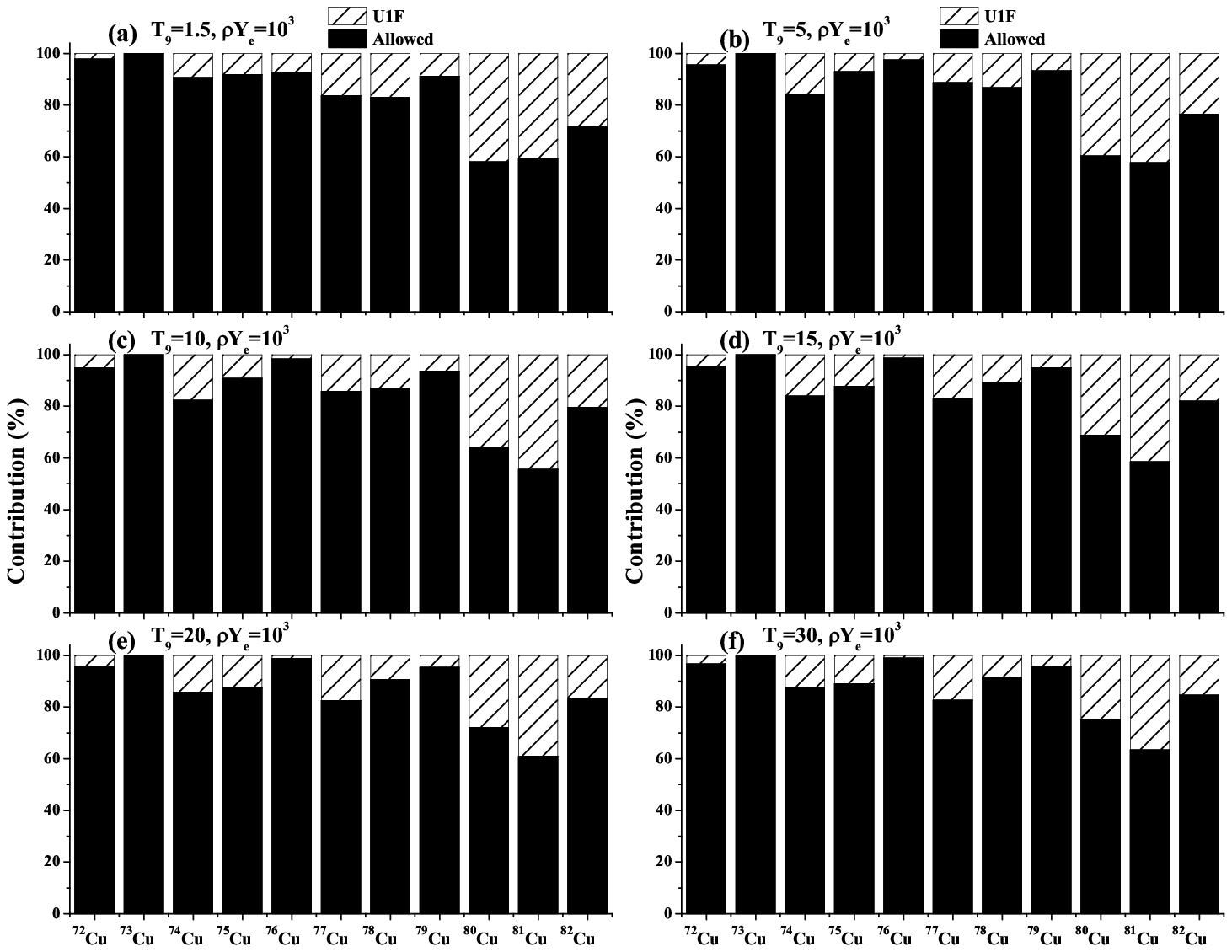}
\caption {Contribution of allowed and U1F rates to total electron
emission rate at low stellar density ($\rho$Y$_{e}$). $\rho$Y$_{e}$
is given in units of g/cm$^{3}$, whereas temperature (T$_{9}$) is
given in units of 10$^{9}$ K.}\label{fig4}
\end{figure*}

\begin{figure*}[h]
\includegraphics[scale=0.6]{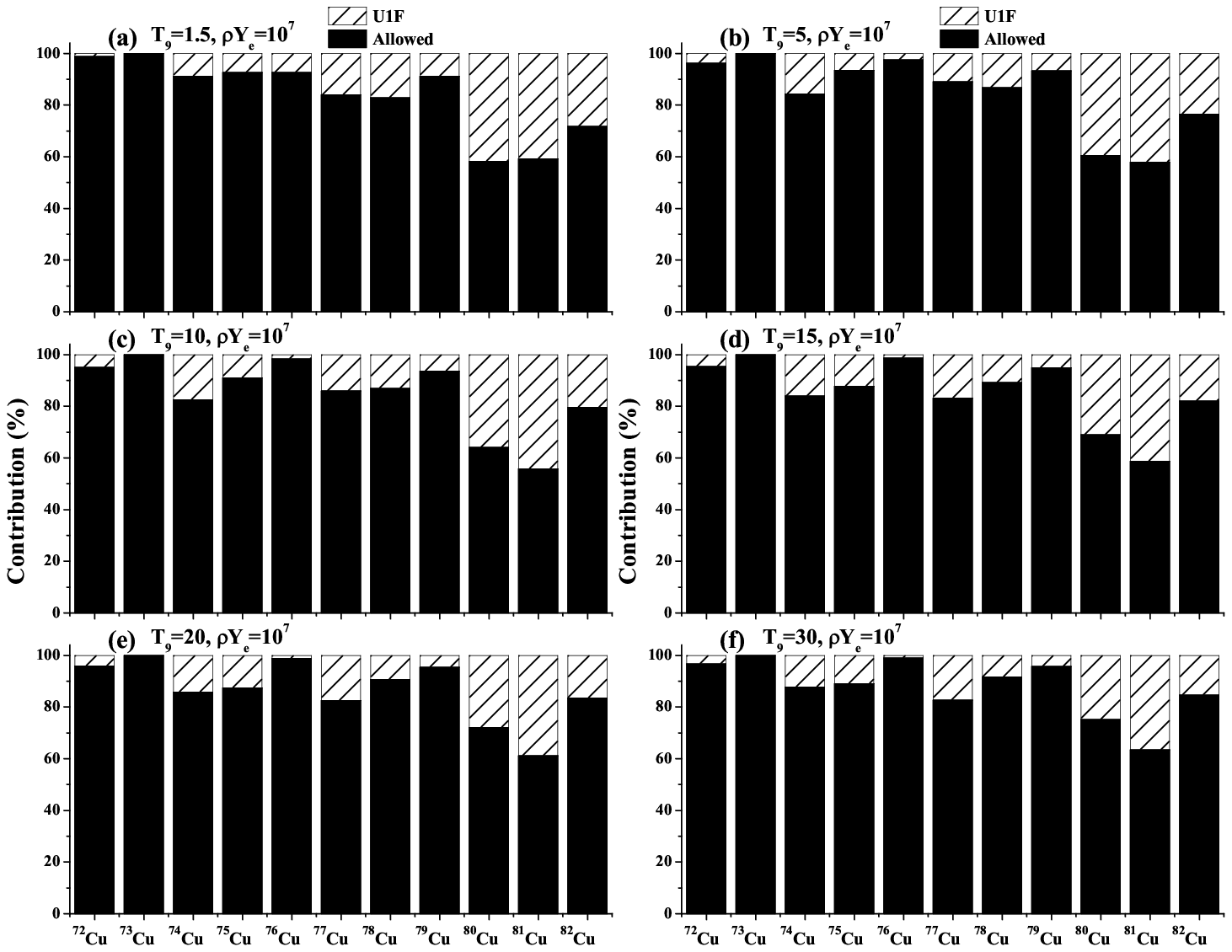}
\caption {Same as Fig.~4, but at intermediate stellar
density.}\label{fig5}
\end{figure*}

\begin{figure*}[h]
\includegraphics[scale=0.6]{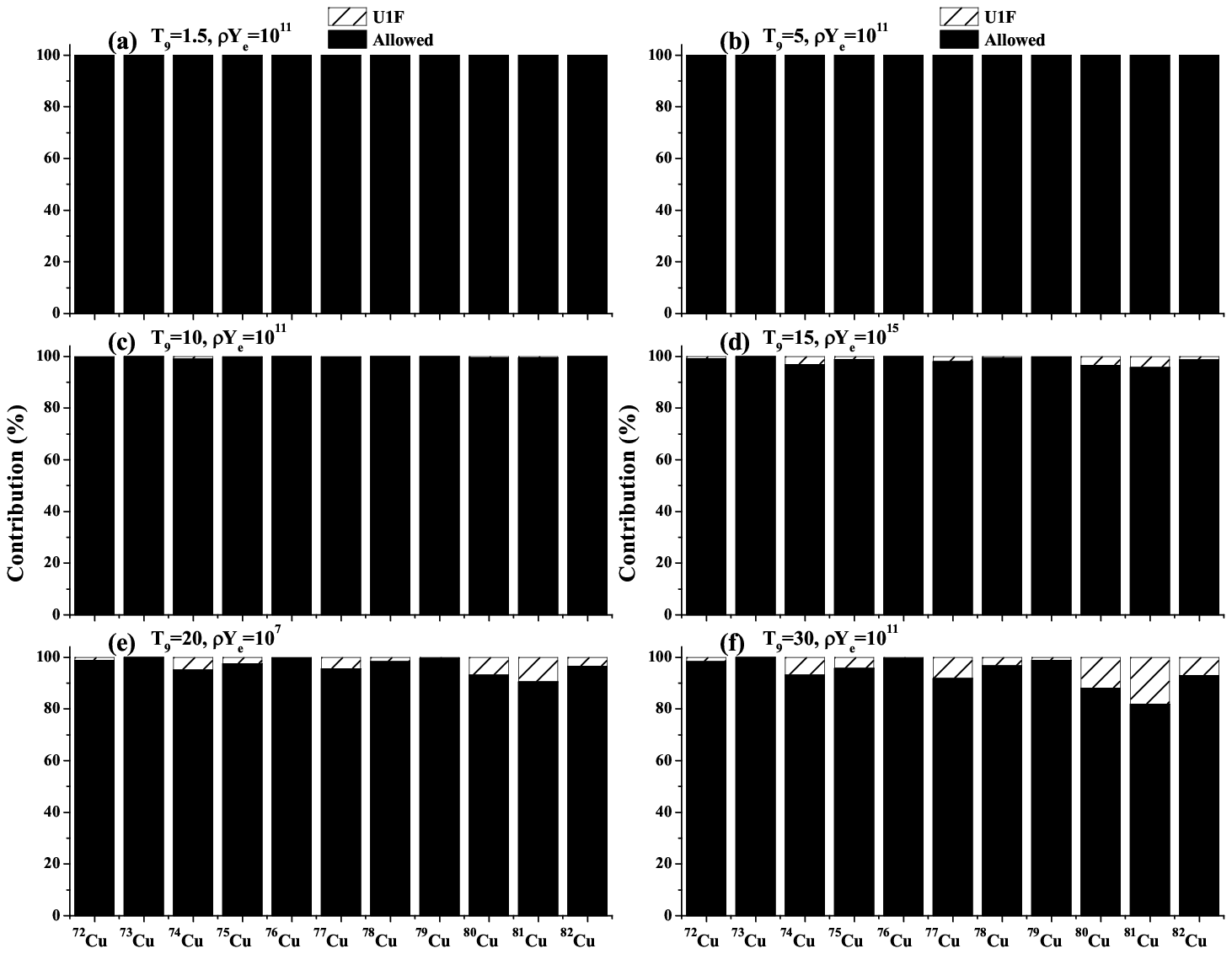}
\caption {Same as Fig.~4, but at high stellar density.}\label{fig6}
\end{figure*}

\clearpage \onecolumn

\begin{table}
\begin{center}
\caption{Comparison of $\beta$-decay half-lives (T$_{1/2}$) of
copper isotopes from different tabulations. In column~2 we list the
T$_{1/2}$ (GT) values from \citep{Mol97}. Columns~3-5 show the
T$_{1/2}$ (GT) calculation of \citep{Pfe02}. Column~6 states the
theoretical (GT + FF) results of \citep{Bor05}. Columns~7 and 8 show
the pn-QRPA (this work) calculated GT and (GT + U1F) results,
respectively. The last three columns give the experimental T$_{1/2}$
values of \citep{Aud12, Aud03} and \citep{Hos10}, respectively}
\label{tab1}

\tiny

\begin{tabular}{c|c|ccc|c|cc|c|c|c}
\hline \hline
\multicolumn{10}{c}{$\beta$-decay half-lives T$_{1/2}$(s)}\\
 Nuclei  & \multicolumn{1}{c|} {M\"{o}ller et al.} &
\multicolumn{3}{c|}{Pfeiffer et al. (GT)} & Borzov et al. &
 \multicolumn{2}{c|}{This work (pn-QRPA)} &   \multicolumn{2}{c|} {Audi et al. (Exp)} & Hosmer et al.\\

 & (GT)& KHF & QRPA-1 &  QRPA-2 & (GT+FF) & (GT) & (GT+U1F) & (2012) & (2003) & (Exp) \\
 \cline{2-11}
    $^{72}$Cu & 6.648 & -     & -     & -     & 7.420 & 7.093 & 6.971 & 6.630 & 6.600  & - \\
    $^{73}$Cu & 3.879 & 1.936 & 4.092 & 2.726 & 4.440 & 4.987 & 4.890 & 4.200 & 4.200  & -  \\
    $^{74}$Cu & 1.657 & 0.393 & 1.308 & 0.975 & 1.480 & 1.905 & 1.759 & 1.630 & 1.594   & - \\
    $^{75}$Cu & 1.381 & 0.458 & 1.345 & 0.844 & 1.320 & 1.530 & 1.409 & 1.224 & 1.224  & - \\
    $^{76}$Cu & 0.830 & 0.153 & 0.657 & 0.428 & 0.736 & 0.792 & 0.685 & 0.638 & 0.641 & 0.599\\
    $^{77}$Cu & 0.833 & 0.155 & 0.764 & 0.405 & 0.590 & 0.601 & 0.490 & 0.468 & 0.469 & 0.466\\
    $^{78}$Cu & 0.480 & 0.077 & 0.351 & 0.268 & 0.424 & 0.461 & 0.372 & 0.335 & 0.342 & 0.335 \\
    $^{79}$Cu & 0.430 & 0.076 & 0.358 & 0.212 & 0.283 & 0.262 & 0.231 & 0.220 & 0.188 & 0.257 \\
    $^{80}$Cu & 0.133 & 0.026 & 0.097 & 0.149 & 0.157 & 0.389 & 0.222 & 0.210 & -     & 0.170\\
    $^{81}$Cu & 0.102 & 0.028 & 0.090 & 0.170 & 0.101 & 0.070 & 0.055 & -     & -    & -\\
    $^{82}$Cu & 0.047 & 0.018 & 0.040 & 0.040 & 0.084 & 0.069 & 0.051 & -     & -    & -\\

\hline \hline
\end{tabular}
\end{center}
\end{table}
\begin{table}
\begin{center}
\caption{Comparison of pn-QRPA calculated $\beta$-decay half-lives
(T$_{1/2}$) for $^{79}$Cu with LSSM  \citep{Zhi13} and measured
data}\label{tab2}

\begin{tabular}{c|c|c|cc|c}
\hline
\multicolumn{6}{c}{$\beta$-decay half-lives T$_{1/2}$(s)}\\
 Nucleus & Zhi et al.  & This work (pn-QRPA) &
\multicolumn{2}{c|} {Audi et al. (Exp)} & Hosmer et al.\\

 &(GT+FF) & (GT+U1F) & (2012) & (2003) & (Exp) \\
 \cline{2-6}

    $^{79}$Cu & 0.270 & 0.231 & 0.220 & 0.188 & 0.257 \\

\hline
\end{tabular}
\end{center}
\end{table}
\begin{table}
\begin{center}
\caption{Comparison of calculated allowed EE rates with and without
insertion of experimental data. $\lambda_{exp}$ and $\lambda_{th}$
show the calculated electron emission (EE) rates with and without
insertion of experimental data, respectively. The first column shows
the stellar densities ($\rho$Y$_{e}$) (in units of gcm$^{-3}$).
T$_{9}$ are given in units of 10$^{9}$ K. The calculated EE rates
are given in units of s$^{-1}$}\label{Table 3} \tiny
\begin{tabular}{|c|c|cc|cc|cc|cc|cc|}
$\rho$$\it Y_{e}$ & T$_{9}$  & \multicolumn{2}{c|}{$^{72}$Cu}& \multicolumn{2}{c|}{$^{73}$Cu}& \multicolumn{2}{c|}{$^{74}$Cu} &\multicolumn{2}{c|}{$^{75}$Cu}&\multicolumn{2}{c|}{$^{77}$Cu}\\
\cline{3-12} & &{$\lambda_{exp}$} &
{$\lambda_{th}$}&{$\lambda_{exp}$} &
{$\lambda_{th}$}&{$\lambda_{exp}$} & {$\lambda_{th}$} &
{$\lambda_{exp}$} & {$\lambda_{th}$} & {$\lambda_{exp}$} &
{$\lambda_{th}$}\\
\hline
     & 1     & 9.91E-02 & 3.12E-01 & 1.64E+00 & 4.26E+00 & 3.84E-01 & 2.11E+00 & 4.46E-01 & 2.00E+00 & 1.18E+00 & 1.77E+00 \\
     & 1.5   & 1.03E-01 & 2.94E-01 & 5.09E+00 & 7.55E+00 & 3.97E-01 & 1.78E+00 & 4.31E-01 & 1.79E+00 & 1.29E+00 & 1.86E+00 \\
     & 2     & 1.09E-01 & 2.79E-01 & 8.75E+00 & 1.10E+01 & 4.08E-01 & 1.56E+00 & 4.19E-01 & 1.69E+00 & 1.45E+00 & 1.99E+00 \\
     & 3     & 1.23E-01 & 2.61E-01 & 1.45E+01 & 1.63E+01 & 4.33E-01 & 1.29E+00 & 4.24E-01 & 1.58E+00 & 1.76E+00 & 2.24E+00 \\
     & 5     & 1.52E-01 & 2.47E-01 & 2.18E+01 & 2.32E+01 & 4.84E-01 & 1.05E+00 & 4.86E-01 & 1.50E+00 & 2.14E+00 & 2.54E+00 \\
$\sim$10--10$^{6}$    & 10    & 2.25E-01 & 2.74E-01 & 3.10E+01 & 3.18E+01 & 6.85E-01 & 9.82E-01 & 7.40E-01 & 1.44E+00 & 2.94E+00 & 3.19E+00 \\
     & 15    & 4.26E-01 & 4.56E-01 & 3.50E+01 & 3.55E+01 & 1.35E+00 & 1.54E+00 & 1.40E+00 & 1.83E+00 & 5.81E+00 & 5.98E+00 \\
     & 20    & 7.74E-01 & 7.96E-01 & 3.89E+01 & 3.92E+01 & 2.52E+00 & 2.67E+00 & 2.56E+00 & 2.81E+00 & 1.07E+01 & 1.08E+01 \\
     & 25    & 1.19E+00 & 1.20E+00 & 4.37E+01 & 4.40E+01 & 3.94E+00 & 4.05E+00 & 4.06E+00 & 4.21E+00 & 1.63E+01 & 1.64E+01 \\
     & 30    & 1.58E+00 & 1.60E+00 & 4.90E+01 & 4.91E+01 & 5.32E+00 & 5.41E+00 & 5.66E+00 & 5.74E+00 & 2.15E+01 & 2.16E+01 \\
\hline
    & 10    & 2.81E-11    &  3.60E-11    & 4.19E-09 & 4.30E-09 & 2.56E-10 & 6.55E-10 & 4.41E-10 & 7.16E-10 & 5.81E-09 & 6.37E-09 \\
    & 15    & 1.12E-07 & 1.21E-07 & 9.51E-06 & 9.64E-06 & 6.58E-07 & 8.51E-07 & 1.01E-06 & 1.19E-06 & 6.61E-06 & 6.85E-06 \\
$\sim$10$^{7}$--10$^{11}$   & 20    & 1.01E-05 & 1.04E-05 & 5.47E-04 & 5.51E-04 & 4.88E-05 & 5.32E-05 & 7.26E-05 & 7.62E-05 & 3.61E-04 & 3.66E-04 \\
    & 25    & 1.70E-04 & 1.73E-04 & 6.95E-03 & 6.98E-03 & 7.41E-04 & 7.71E-04 & 1.08E-03 & 1.10E-03 & 4.66E-03 & 4.69E-03 \\
   & 30    & 1.17E-03 & 1.18E-03 & 4.04E-02 & 4.05E-02 & 4.81E-03 & 4.91E-03 & 6.87E-03 & 6.92E-03 & 2.70E-02 & 2.71E-02 \\
\hline
\end{tabular}
\end{center}
\end{table}


\begin{table*}
\centering \scriptsize \caption{Calculated allowed and unique
first-forbidden (U1F) electron emission rates on $^{72-75}$Cu
isotopes for different selected densities and temperatures in
stellar matter. The first column shows the stellar densities
($\rho$Y$_{e}$) (in units of gcm$^{-3}$). T$_{9}$ are given in units
of 10$^{9}$ K. The calculated emission rates are tabulated in
logarithmic (to base 10) scale in units of s$^{-1}$}\label{Table 4}
    \begin{tabular}{|c|c|cc|cc|cc|cc|}

$\rho$$\it Y_{e}$ & T$_{9}$ & \multicolumn{2}{c|}{$^{72}$Cu}& \multicolumn{2}{c|}{$^{73}$Cu}& \multicolumn{2}{c|}{$^{74}$Cu} & \multicolumn{2}{c|}{$^{75}$Cu}\\
\cline{3-10} & &Allowed & U1F& Allowed & U1F&Allowed & U1F & Allowed & U1F\\
\hline
10$^{1}$ & 1     & -1.00 & -2.74 & 0.21  & -2.56 & -0.42 & -1.47 & -0.35 & -1.41 \\
10$^{1}$ & 3     & -0.91 & -2.34 & 1.16  & -2.59 & -0.36 & -1.17 & -0.37 & -1.44 \\
10$^{1}$ & 5     & -0.82 & -2.15 & 1.34  & -2.58 & -0.32 & -1.03 & -0.31 & -1.43 \\
10$^{1}$ & 10    & -0.65 & -1.92 & 1.49  & -2.26 & -0.16 & -0.83 & -0.13 & -1.13 \\
10$^{1}$ & 30    & 0.20  & -1.25 & 1.69  & -1.34 & 0.73  & -0.13 & 0.75  & -0.15 \\
\hline
10$^{5}$ & 1     & -1.00 & -2.74 & 0.21  & -2.56 & -0.42 & -1.47 & -0.35 & -1.41 \\
10$^{5}$ & 3     & -0.91 & -2.34 & 1.16  & -2.59 & -0.36 & -1.18 & -0.37 & -1.44 \\
10$^{5}$ & 5     & -0.82 & -2.15 & 1.34  & -2.58 & -0.32 & -1.03 & -0.31 & -1.43 \\
10$^{5}$ & 10    & -0.65 & -1.92 & 1.49  & -2.26 & -0.16 & -0.83 & -0.13 & -1.13 \\
10$^{5}$ & 30    & -0.20  & -1.25 & 1.69  & -1.34 & 0.73  & -0.13 & 0.75  & -0.15 \\
\hline
10$^{10}$ & 1    & -30.58 & -50.40 & -24.09 & -50.09 & -23.74 & -41.29 & -21.76 & -44.07 \\
10$^{10}$ & 3    & -11.68 & -18.36 & -8.65 & -18.72 & -9.38 & -14.97 & -8.65 & -16.23 \\
10$^{10}$ & 5    & -7.54 & -11.76 & -5.00 & -12.25 & -6.09 & -9.44 & -5.71 & -10.39 \\
10$^{10}$ & 10   & -4.07 & -6.60 & -1.87 & -6.94 & -3.19 & -5.02 & -3.05 & -5.51 \\
10$^{10}$ & 30   & -0.67 & -2.38 & 0.85  & -2.46 & -0.09 & -1.15 & 0.01  & -1.21 \\
\hline

\end{tabular}
\end{table*}

\begin{table*}
\centering \scriptsize \caption{Same as Table~4 but for $^{76-79}$Cu
isotopes}\label{Table 5}
    \begin{tabular}{|c|c|cc|cc|cc|cc|}

$\rho$$\it Y_{e}$ & T$_{9}$ & \multicolumn{2}{c|}{$^{76}$Cu}& \multicolumn{2}{c|}{$^{77}$Cu}& \multicolumn{2}{c|}{$^{78}$Cu} & \multicolumn{2}{c|}{$^{79}$Cu}\\
\cline{3-10} & &Allowed & U1F& Allowed & U1F&Allowed & U1F & Allowed & U1F\\
\hline
10$^{1}$ & 1     & 0.08  & -0.86 & 0.07  & -0.59 & 0.20  & -0.45 & 0.46  & -0.46 \\
10$^{1}$ & 3     & 0.57  & -0.83 & 0.25  & -0.61 & 0.33  & -0.46 & 0.67  & -0.47 \\
10$^{1}$ & 5     & 0.81  & -0.78 & 0.33  & -0.56 & 0.44  & -0.38 & 0.72  & -0.41 \\
10$^{1}$ & 10    & 1.15  & -0.62 & 0.47  & -0.31 & 0.64  & -0.18 & 0.91  & -0.26 \\
10$^{1}$ & 30    & 2.09  & 0.12  & 1.33  & 0.66  & 1.54  & 0.50  & 2.07  & 0.72 \\
\hline
10$^{5}$ & 1     & 0.08  & -0.86 & 0.07  & -0.59 & 0.19  & -0.45 & 0.46  & -0.47 \\
10$^{5}$ & 3     & 0.57  & -0.83 & 0.25  & -0.61 & 0.33  & -0.46 & 0.67  & -0.47 \\
10$^{5}$ & 5     & 0.81  & -0.78 & 0.33  & -0.56 & 0.44  & -0.38 & 0.72  & -0.41 \\
10$^{5}$ & 10    & 1.15  & -0.62 & 0.47  & -0.31 & 0.64  & -0.18 & 0.91  & -0.26 \\
10$^{5}$ & 30    & 2.09  & 0.12  & 1.33  & 0.66  & 1.54  & 0.50  & 2.07  & 0.72 \\
\hline
10$^{10}$ & 1    & -18.38 & -40.06 & -14.22 & -37.24 & -12.96 & -35.82 & -11.67 & -34.89 \\
10$^{10}$ & 3    & -6.47 & -14.61 & -6.00 & -13.62 & -5.74 & -13.12 & -5.16 & -12.83 \\
10$^{10}$ & 5    & -3.76 & -9.21 & -3.96 & -8.55 & -3.69 & -8.20 & -3.11 & -8.06 \\
10$^{10}$ & 10   & -1.35 & -4.82 & -1.99 & -4.30 & -1.64 & -4.12 & -1.14 & -4.12 \\
10$^{10}$ & 30   & 1.36  & -0.91 & 0.63  & -0.32 & 0.84  & -0.47 & 1.45  & -0.24 \\
\hline

\end{tabular}
\end{table*}

\begin{table*}
\centering \scriptsize \caption{Same as Table~4 but for $^{80-82}$Cu
isotopes}\label{Table 6}
    \begin{tabular}{|c|c|cc|cc|cc|}

$\rho$$\it Y_{e}$ & T$_{9}$ & \multicolumn{2}{c|}{$^{80}$Cu}& \multicolumn{2}{c|}{$^{81}$Cu}& \multicolumn{2}{c|}{$^{82}$Cu} \\
\cline{3-8} & &Allowed & U1F& Allowed & U1F&Allowed & U1F\\
\hline
10$^{1}$ & 1     & 0.23  & 0.10  & 0.59  & 0.43  & 0.93  & 0.54 \\
10$^{1}$ & 3     & 0.26  & 0.10  & 0.60  & 0.44  & 0.98  & 0.53 \\
10$^{1}$ & 5     & 0.34  & 0.15  & 0.62  & 0.48  & 1.08  & 0.57 \\
10$^{1}$ & 10    & 0.55  & 0.30  & 0.74  & 0.64  & 1.33  & 0.74 \\
10$^{1}$ & 30    & 1.45  & 0.97  & 1.87  & 1.63  & 2.21  & 1.47 \\
\hline
10$^{5}$ & 1     & 0.23  & 0.10  & 0.59  & 0.43  & 0.93  & 0.54 \\
10$^{5}$ & 3     & 0.26  & 0.10  & 0.60  & 0.44  & 0.98  & 0.53 \\
10$^{5}$ & 5     & 0.34  & 0.15  & 0.62  & 0.48  & 1.08  & 0.57 \\
10$^{5}$ & 10    & 0.55  & 0.30  & 0.74  & 0.64  & 1.33  & 0.74 \\
10$^{5}$ & 30    & 1.45  & 0.97  & 1.87  & 1.63  & 2.21  & 1.47 \\
\hline
10$^{10}$ & 1     & -8.52 & -30.05 & -3.22 & -23.66 & -3.23 & -22.37 \\
10$^{10}$ & 3     & -4.21 & -11.00 & -2.61 & -8.89 & -2.62 & -8.45 \\
10$^{10}$ & 5     & -2.94 & -6.81 & -2.17 & -5.53 & -1.92 & -5.28 \\
10$^{10}$ & 10    & -1.48 & -3.28 & -1.21 & -2.55 & -0.57 & -2.41 \\
10$^{10}$ & 30    & 0.81  & 0.07  & 1.26  & 0.80  & 1.60  & 0.64 \\
\hline
\end{tabular}
\end{table*}
\begin{table*}
\centering \scriptsize \caption{Calculated allowed (GT) and unique
first-forbidden (U1F) electron emission (EE) rates on $^{72-75}$Cu
isotopes for different selected densities and temperatures in
stellar matter. The first column shows the stellar densities
($\rho$Y$_{e}$) (in units of gcm$^{-3}$). T$_{9}$ are given in units
of 10$^{9}$ K. The calculated emission rates are tabulated in
logarithmic (to base 10) scale in units of s$^{-1}$. The third
column in each box gives the percentage contribution of allowed GT
to total EE rates}\label{Table 7}
    \begin{tabular}{|c|c|ccc|ccc|ccc|ccc|}
\hline
$\rho$$\it Y_{e}$ & T$_{9}$ & \multicolumn{3}{c|}{$^{72}$Cu}& \multicolumn{3}{c|}{$^{73}$Cu}& \multicolumn{3}{c|}{$^{74}$Cu} & \multicolumn{3}{c|}{$^{75}$Cu}\\
\cline{3-14} & & GT & U1F& \%(GT)& GT & U1F& \%(GT)&GT & U1F& \%(GT) & GT & U1F& \%(GT)\\
\hline
       & 1.5   & -0.99 & -2.65 & 97.85 & 0.71  & -2.57 & 99.95 & -0.40 & -1.39 & 90.68 & -0.37 & -1.41 & 91.78 \\
       & 2     & -0.96 & -2.53 & 97.36 & 0.94  & -2.57 & 99.97 & -0.39 & -1.30 & 89.16 & -0.38 & -1.42 & 91.71 \\
       & 3     & -0.91 & -2.34 & 96.44 & 1.16  & -2.59 & 99.98 & -0.36 & -1.17 & 86.56 & -0.37 & -1.44 & 92.09 \\
       & 5     & -0.82 & -2.15 & 95.56 & 1.34  & -2.58 & 99.99 & -0.32 & -1.03 & 83.93 & -0.31 & -1.43 & 92.93 \\
10$^{3}$ & 10    & -0.65 & -1.92 & 94.95 & 1.49  & -2.26 & 99.98 & -0.16 & -0.83 & 82.35 & -0.13 & -1.13 & 90.79 \\
       & 15    & -0.37 & -1.68 & 95.30 & 1.54  & -1.86 & 99.96 & 0.13  & -0.59 & 83.96 & 0.15  & -0.70 & 87.62 \\
       & 20    & -0.11 & -1.48 & 95.92 & 1.59  & -1.60 & 99.94 & 0.40  & -0.38 & 85.77 & 0.41  & -0.43 & 87.32 \\
       & 25    & 0.07  & -1.35 & 96.34 & 1.64  & -1.44 & 99.92 & 0.60  & -0.23 & 86.93 & 0.61  & -0.26 & 88.11 \\
       & 30    & 0.20  & -1.25 & 96.60 & 1.69  & -1.34 & 99.91 & 0.73  & -0.13 & 87.65 & 0.75  & -0.15 & 88.93 \\
       \hline
       & 1.5   & -1.02 & -2.93 & 98.80 & 0.69  & -2.79 & 99.97 & -0.42 & -1.44 & 91.17 & -0.38 & -1.48 & 92.58 \\
       & 2     & -0.99 & -2.76 & 98.33 & 0.92  & -2.78 & 99.98 & -0.41 & -1.35 & 89.62 & -0.39 & -1.49 & 92.51 \\
       & 3     & -0.94 & -2.50 & 97.37 & 1.14  & -2.75 & 99.99 & -0.38 & -1.21 & 86.98 & -0.39 & -1.50 & 92.77 \\
10$^{7}$& 5     & -0.84 & -2.23 & 96.14 & 1.33  & -2.67 & 99.99 & -0.33 & -1.06 & 84.24 & -0.32 & -1.47 & 93.35 \\
       & 10    & -0.65 & -1.94 & 95.08 & 1.49  & -2.28 & 99.98 & -0.17 & -0.84 & 82.49 & -0.14 & -1.14 & 90.95 \\
       & 15    & -0.37 & -1.69 & 95.34 & 1.54  & -1.87 & 99.96 & 0.13  & -0.60 & 84.03 & 0.14  & -0.71 & 87.70 \\
       & 20    & -0.11 & -1.49 & 95.93 & 1.59  & -1.61 & 99.94 & 0.40  & -0.38 & 85.82 & 0.41  & -0.43 & 87.37 \\
       & 25    & 0.07  & -1.35 & 96.35 & 1.64  & -1.45 & 99.92 & 0.59  & -0.23 & 86.96 & 0.61  & -0.26 & 88.11 \\
       & 30    & 0.20  & -1.25 & 96.60 & 1.69  & -1.34 & 99.91 & 0.73  & -0.13 & 87.65 & 0.75  & -0.15 & 88.95 \\
       \hline
       & 1.5   & -64.33 & -77.58 & 100.00 & -59.62 & -77.58 & 100.00 & -59.43 & -71.35 & 100.00 & -55.80 & -73.35 & 100.00 \\
       & 2     & -48.83 & -58.78 & 100.00 & -45.03 & -58.97 & 100.00 & -45.14 & -53.99 & 100.00 & -42.55 & -55.62 & 100.00 \\
       & 3     & -33.17 & -39.91 & 100.00 & -30.16 & -40.27 & 100.00 & -30.67 & -36.52 & 100.00 & -29.15 & -37.78 & 100.00 \\
10$^{11}$ & 5     & -20.44 & -24.71 & 99.99 & -17.93 & -25.20 & 100.00 & -18.85 & -22.39 & 99.97 & -18.15 & -23.33 & 100.00 \\
       & 10    & -10.55 & -13.12 & 99.73 & -8.38 & -13.46 & 100.00 & -9.59 & -11.53 & 98.87 & -9.36 & -12.02 & 99.79 \\
       & 15    & -6.95 & -9.03 & 99.18 & -5.02 & -9.21 & 99.99 & -6.18 & -7.66 & 96.77 & -6.00 & -7.88 & 98.71 \\
       & 20    & -5.00 & -6.89 & 98.74 & -3.26 & -7.01 & 99.98 & -4.31 & -5.59 & 95.04 & -4.14 & -5.72 & 97.41 \\
       & 25    & -3.77 & -5.57 & 98.45 & -2.16 & -5.67 & 99.97 & -3.13 & -4.31 & 93.84 & -2.97 & -4.40 & 96.42 \\
       & 30    & -2.93 & -4.68 & 98.26 & -1.39 & -4.77 & 99.96 & -2.32 & -3.44 & 93.02 & -2.16 & -3.51 & 95.70 \\
\hline
\end{tabular}
\end{table*}
\begin{table*}
\centering \scriptsize \caption{Same as Table~7 but for $^{76-79}$Cu
isotopes}\label{Table 8}
    \begin{tabular}{|c|c|ccc|ccc|ccc|ccc|}
\hline
$\rho$$\it Y_{e}$ & T$_{9}$ & \multicolumn{3}{c|}{$^{76}$Cu}& \multicolumn{3}{c|}{$^{77}$Cu}& \multicolumn{3}{c|}{$^{78}$Cu} & \multicolumn{3}{c|}{$^{79}$Cu}\\
\cline{3-14} & & GT & U1F& \%(GT)& GT & U1F& \%(GT)&GT & U1F& \%(GT) & GT & U1F& \%(GT)\\
\hline
       & 1.5   & 0.23  & -0.86 & 92.47 & 0.11  & -0.60 & 83.56 & 0.22  & -0.46 & 82.95 & 0.53  & -0.48 & 91.06 \\
       & 2     & 0.37  & -0.86 & 94.37 & 0.16  & -0.61 & 85.34 & 0.26  & -0.47 & 84.39 & 0.59  & -0.49 & 92.27 \\
       & 3     & 0.57  & -0.83 & 96.21 & 0.25  & -0.61 & 87.70 & 0.33  & -0.46 & 86.18 & 0.67  & -0.47 & 93.20 \\
       & 5     & 0.81  & -0.78 & 97.48 & 0.33  & -0.56 & 88.70 & 0.44  & -0.38 & 86.70 & 0.72  & -0.41 & 93.22 \\
10$^{3}$ & 10    & 1.15  & -0.62 & 98.33 & 0.47  & -0.31 & 85.77 & 0.64  & -0.18 & 86.93 & 0.91  & -0.26 & 93.61 \\
       & 15    & 1.49  & -0.38 & 98.68 & 0.76  & 0.07  & 83.04 & 0.95  & 0.03  & 89.18 & 1.34  & 0.08  & 94.82 \\
       & 20    & 1.78  & -0.16 & 98.84 & 1.03  & 0.36  & 82.45 & 1.22  & 0.24  & 90.60 & 1.69  & 0.38  & 95.30 \\
       & 25    & 1.96  & 0.01  & 98.91 & 1.21  & 0.54  & 82.49 & 1.41  & 0.39  & 91.25 & 1.92  & 0.59  & 95.53 \\
       & 30    & 2.09  & 0.12  & 98.94 & 1.33  & 0.66  & 82.62 & 1.54  & 0.50  & 91.57 & 2.07  & 0.72  & 95.69 \\
\hline
       & 1.5   & 0.20  & -0.89 & 92.51 & 0.10  & -0.62 & 84.00 & 0.20  & -0.48 & 82.78 & 0.52  & -0.50 & 91.19 \\
       & 2     & 0.34  & -0.89 & 94.44 & 0.15  & -0.63 & 85.77 & 0.24  & -0.49 & 84.30 & 0.58  & -0.51 & 92.39 \\
       & 3     & 0.55  & -0.87 & 96.30 & 0.24  & -0.63 & 88.07 & 0.31  & -0.48 & 86.18 & 0.65  & -0.49 & 93.33 \\
10$^{7}$ & 5     & 0.80  & -0.80 & 97.54 & 0.33  & -0.58 & 88.98 & 0.42  & -0.39 & 86.75 & 0.72  & -0.43 & 93.32 \\
       & 10    & 1.14  & -0.63 & 98.35 & 0.47  & -0.32 & 85.88 & 0.63  & -0.19 & 86.96 & 0.90  & -0.26 & 93.64 \\
       & 15    & 1.49  & -0.39 & 98.69 & 0.76  & 0.07  & 83.11 & 0.94  & 0.03  & 89.22 & 1.34  & 0.08  & 94.84 \\
       & 20    & 1.77  & -0.16 & 98.84 & 1.03  & 0.36  & 82.49 & 1.22  & 0.24  & 90.62 & 1.69  & 0.38  & 95.30 \\
       & 25    & 1.96  & 0.01  & 98.91 & 1.21  & 0.54  & 82.52 & 1.41  & 0.39  & 91.25 & 1.92  & 0.59  & 95.53 \\
       & 30    & 2.09  & 0.12  & 98.94 & 1.33  & 0.66  & 82.62 & 1.54  & 0.50  & 91.59 & 2.06  & 0.72  & 95.69 \\
\hline
       & 1.5   & -53.35 & -70.57 & 100.00 & -49.35 & -68.67 & 100.00 & -48.13 & -67.70 & 100.00 & -43.90 & -67.12 & 100.00 \\
       & 2     & -40.38 & -53.43 & 100.00 & -37.71 & -51.99 & 100.00 & -36.63 & -51.25 & 100.00 & -33.38 & -50.83 & 100.00 \\
       & 3     & -27.20 & -36.16 & 100.00 & -25.86 & -35.17 & 100.00 & -24.97 & -34.67 & 100.00 & -22.69 & -34.38 & 100.00 \\
10$^{11}$ & 5     & -16.36 & -22.16 & 100.00 & -16.09 & -21.50 & 100.00 & -15.36 & -21.14 & 100.00 & -13.89 & -21.01 & 100.00 \\
       & 10    & -7.74 & -11.34 & 99.98 & -8.24 & -10.82 & 99.74 & -7.67 & -10.64 & 99.89 & -6.86 & -10.64 & 99.98 \\
       & 15    & -4.50 & -7.46 & 99.89 & -5.18 & -6.87 & 98.01 & -4.74 & -6.90 & 99.30 & -4.04 & -6.80 & 99.83 \\
       & 20    & -2.72 & -5.37 & 99.78 & -3.44 & -4.77 & 95.55 & -3.10 & -4.88 & 98.37 & -2.38 & -4.70 & 99.52 \\
       & 25    & -1.60 & -4.08 & 99.67 & -2.33 & -3.48 & 93.40 & -2.04 & -3.62 & 97.43 & -1.32 & -3.40 & 99.18 \\
       & 30    & -0.83 & -3.20 & 99.58 & -1.57 & -2.61 & 91.71 & -1.31 & -2.76 & 96.62 & -0.59 & -2.52 & 98.84 \\
\hline
\end{tabular}
\end{table*}
\begin{table*}
\centering \scriptsize \caption{Same as Table~7 but for $^{80-82}$Cu
isotopes}\label{Table 9}
    \begin{tabular}{|c|c|ccc|ccc|ccc|}
\hline
$\rho$$\it Y_{e}$ & T$_{9}$ & \multicolumn{3}{c|}{$^{80}$Cu}& \multicolumn{3}{c|}{$^{81}$Cu}& \multicolumn{3}{c|}{$^{82}$Cu}\\
\cline{3-11} & & GT & U1F& \%(GT)& GT & U1F& \%(GT)&GT & U1F& \%(GT)\\
\hline
       & 1.5   & 0.23  & 0.09  & 57.99 & 0.59  & 0.43  & 59.16 & 0.94  & 0.54  & 71.62 \\
       & 2     & 0.23  & 0.08  & 58.61 & 0.60  & 0.43  & 59.22 & 0.95  & 0.53  & 72.41 \\
       & 3     & 0.26  & 0.10  & 59.50 & 0.60  & 0.44  & 59.11 & 0.98  & 0.53  & 73.99 \\
       & 5     & 0.34  & 0.15  & 60.33 & 0.62  & 0.48  & 57.71 & 1.08  & 0.57  & 76.39 \\
10$^{3}$ & 10    & 0.55  & 0.30  & 64.06 & 0.74  & 0.64  & 55.62 & 1.33  & 0.74  & 79.48 \\
       & 15    & 0.84  & 0.50  & 68.78 & 1.14  & 0.99  & 58.55 & 1.63  & 0.97  & 81.95 \\
       & 20    & 1.11  & 0.70  & 72.08 & 1.49  & 1.30  & 61.04 & 1.90  & 1.19  & 83.49 \\
       & 25    & 1.31  & 0.86  & 73.94 & 1.72  & 1.50  & 62.51 & 2.08  & 1.36  & 84.18 \\
       & 30    & 1.45  & 0.97  & 75.04 & 1.87  & 1.63  & 63.37 & 2.21  & 1.47  & 84.54 \\
\hline
       & 1.5   & 0.22  & 0.08  & 58.16 & 0.59  & 0.42  & 59.16 & 0.93  & 0.53  & 71.67 \\
       & 2     & 0.23  & 0.07  & 58.83 & 0.59  & 0.42  & 59.22 & 0.94  & 0.52  & 72.45 \\
       & 3     & 0.25  & 0.08  & 59.66 & 0.59  & 0.43  & 59.16 & 0.98  & 0.52  & 74.08 \\
10$^{7}$ & 5     & 0.33  & 0.14  & 60.49 & 0.61  & 0.47  & 57.82 & 1.08  & 0.56  & 76.48 \\
       & 10    & 0.55  & 0.30  & 64.11 & 0.74  & 0.64  & 55.67 & 1.32  & 0.73  & 79.51 \\
       & 15    & 0.84  & 0.49  & 68.88 & 1.14  & 0.99  & 58.61 & 1.63  & 0.97  & 81.98 \\
       & 20    & 1.11  & 0.70  & 72.13 & 1.49  & 1.30  & 61.09 & 1.90  & 1.19  & 83.49 \\
       & 25    & 1.31  & 0.86  & 73.99 & 1.72  & 1.50  & 62.51 & 2.08  & 1.36  & 84.21 \\
       & 30    & 1.45  & 0.97  & 75.08 & 1.87  & 1.63  & 63.42 & 2.21  & 1.47  & 84.54 \\
\hline
       & 1.5   & -43.91 & -63.77 & 100.00 & -36.09 & -59.58 & 100.00 & -35.92 & -58.65 & 100.00 \\
       & 2     & -33.72 & -48.23 & 100.00 & -28.01 & -45.09 & 100.00 & -27.81 & -44.40 & 100.00 \\
       & 3     & -23.30 & -32.54 & 100.00 & -19.76 & -30.44 & 100.00 & -19.40 & -30.00 & 100.00 \\
10$^{11}$& 5     & -14.66 & -19.76 & 100.00 & -12.85 & -18.48 & 100.00 & -12.16 & -18.22 & 100.00 \\
       & 10    & -7.58 & -9.80 & 99.40 & -6.87 & -9.07 & 99.36 & -6.08 & -8.93 & 99.86 \\
       & 15    & -4.78 & -6.22 & 96.53 & -4.15 & -5.49 & 95.68 & -3.63 & -5.50 & 98.66 \\
       & 20    & -3.15 & -4.28 & 93.05 & -2.55 & -3.52 & 90.40 & -2.18 & -3.62 & 96.54 \\
       & 25    & -2.09 & -3.05 & 90.18 & -1.52 & -2.29 & 85.60 & -1.20 & -2.43 & 94.49 \\
       & 30    & -1.35 & -2.21 & 87.99 & -0.81 & -1.46 & 81.85 & -0.51 & -1.62 & 92.84 \\
\hline

\end{tabular}
\end{table*}

\end{document}